\documentclass[utf8]{FrontiersinHarvard} % for articles in journals using the Harvard Referencing Style (Author-Date), for Frontiers Reference Styles by Journal: https://zendesk.frontiersin.org/hc/en-us/articles/360017860337-Frontiers-Reference-Styles-by-Journal
\usepackage{url,hyperref,lineno,microtype,subcaption}
\usepackage[onehalfspacing]{setspace}
\newcommand{\red}[1]{\textcolor{red}{#1}}
\usepackage{bookmark}
\usepackage{multirow}
\usepackage{graphicx}
\usepackage{booktabs}
\usepackage{float}
\usepackage{mathtools}
\usepackage{amsmath}
\usepackage{amsfonts}
\usepackage{bm}
\usepackage{amssymb}
\usepackage{stackengine}
\usepackage{algorithm}
\usepackage{algpseudocode}
\DeclareMathOperator{\E}{\mathbb{E}}

\def\keyFont{\fontsize{8}{11}\helveticabold }
\def\firstAuthorLast{Xia {et~al.}} %use et al only if is more than 1 author
\def\Authors{Tian Xia\,$^{1,*}$, Pedro Sanchez\,$^{1}$, Chen Qin\,$^{1}$ and Sotirios A. Tsaftaris\,$^{1,2}$}
\def\Address{$^{1}$School of Engineering, University of Edinburgh, UK\\
$^{2}$The Alan Turing Institute, London, UK }
\def\corrAuthor{Corresponding Author}

\def\corrEmail{tian.xia@ed.ac.uk}

\begin{document}
\onecolumn
\firstpage{1}

\title[Adversarial Counterfactual Augmentation: Application in AD Classification]{Adversarial Counterfactual Augmentation: Application in Alzheimer's Disease Classification} 

\author[\firstAuthorLast ]{\Authors} %This field will be automatically populated
\address{} %This field will be automatically populated
\correspondance{} %This field will be automatically populated

\extraAuth{}% If there are more than 1 corresponding author, comment this line and uncomment the next one.
%\extraAuth{corresponding Author2 \\ Laboratory X2, Institute X2, Department X2, Organization X2, Street X2, City X2 , State XX2 (only USA, Canada and Australia), Zip Code2, X2 Country X2, email2@uni2.edu}

\maketitle

\begin{abstract}

Due to the limited availability of medical data, deep learning approaches for medical image analysis tend to generalise poorly to unseen data. Augmenting data during training with random transformations has been shown to help and became a ubiquitous technique for training neural networks. Here, we propose a novel adversarial counterfactual augmentation scheme that aims at finding the most \textit{effective} synthesised images to improve downstream tasks, given a pre-trained generative model. Specifically, we construct an adversarial game where we update the input \textit{conditional factor} of the generator and the downstream \textit{classifier} with gradient backpropagation alternatively and iteratively. This can be viewed as finding the `\textit{weakness}' of the classifier and purposely forcing it to \textit{overcome} its weakness via the generative model. To demonstrate the effectiveness of the proposed approach, we validate the method with the classification of Alzheimer's Disease (AD) as a downstream task. The pre-trained generative model synthesises brain images using age as conditional factor. Extensive experiments and ablation studies have been performed to show that the proposed approach improves classification performance and has potential to alleviate spurious correlations and catastrophic forgetting. \\ Code will be released upon acceptance.

\tiny
 \keyFont{ \section{Keywords:}Alzheimer's Disease,  Generative model, Classification, Counterfactuals, Data efficiency} %All article types: you may provide up to 8 keywords; at least 5 are mandatory.
\end{abstract}

\section{Introduction}
 {Deep learning has been playing an increasingly important role in medical image analysis in the past decade, with great success in segmentation, diagnosis, detection, etc~\citep{shen2017deep}. Although deep-learning based models can significantly outperform traditional machine learning methods, they heavily rely on  the large size and quality of training data~\citep{chlap2021review}. In medical image analysis, the availability of large dataset is always an issue, due to high expense of acquiring and labelling medical imaging data~\citep{shorten2019survey}. When only limited training data are available, deep neural networks tend to memorise the data and cannot generalise well to unseen data~\citep{dietterich1995overfitting,srivastava2014dropout}. This is known as \textit{over-fitting}~\citep{dietterich1995overfitting}. To mitigate this issue, data augmentation has become a popular approach. The aim of data augmentation is to generate additional data that can help increase the variation of the training data. }

Conventional data augmentation approaches mainly apply random image transformations, such as cropping, flipping, and rotation etc.\ to the data. Even though such conventional data augmentation techniques are general, they may not transfer well from one task to another~\citep{cubuk2019autoaugment}.  {For instance, color augmentation could prove useful for natural images but may not be suitable for MRI images which are presented in greyscale images~\citep{shorten2019survey}}.
% However, one problem with these conventional data augmentations is that the augmentation strategy that works well for one dataset may not transfer well to another~\cite{cubuk2019autoaugment}.
Furthermore, traditional data augmentation methods may introduce \textit{distribution shift}, i.e., the change of the joint distribution of inputs and outputs, and consequently adversely impact the performance on non-augmented data during inference\footnote{ {An example could be when training and testing brain MRI data are already well-registered, traditional augmentations, e.g. rotation, shift, etc., on the training data will hurt the performance of the trained model on testing data. See Section~\ref{sec: Counterfactual augmentation v.s. conventional augmentation} for more details.}} (i.e., during the application phase of the learned model)~\citep{gong2021keepaugment}.  

Some recently developed approaches learn parameters for data augmentation that can better improve downstream task, e.g. segmentation, detection, diagnosis, etc., performance~\citep{chen2021enhancing,cubuk2019autoaugment,gao2021enabling} or select the hardest augmentation for the target model from a small batch of random augmentations for each traning sample, ~\citep{gong2021maxup}.  However, these approaches still use conventional image transformations
% e.g., cropping, rotation and deformation, 
and do not consider semantic augmentation~\citep{wang2021regularizing}, i.e., creating unseen samples by changing semantic information of images such as changing the background of an object or changing the age of a brain image. Semantic augmentation can complement  traditional techniques and improve the diversity of augmented samples~\citep{wang2021regularizing}.

One way to achieve semantic augmentation is to train a deep generative model to create \textit{counterfactuals}, i.e., synthetic modifications of a sample such that some aspects of the original data remain unchanged~\citep{zhang2020deep,shamsolmoali2021imbalanced,bowles2018gan,oh2021learn,dash2022evaluating}.  {However, these approaches mostly focus on the training stage of generative models and randomly generate samples for data augmentation, without considering which counterfactuals are \textit{more effective} for downstream tasks, i.e. data-efficiency of the generated samples . \cite{wang2021regularizing,li2021metasaug,chen2021sample} proposed to augment the data in the latent space of the target deep neural network, by estimating the covariance matrix of latent features obtained from latent layers of the target deep neural network for each \textit{class} {(e.g., car, horse, tree, etc.) and sampling {directions} from the feature distributions. These directions should be semantic meaningful such that  changing along one direction can manipulate one property of the image, e.g. color of a car.} However, there is no guarantee that the found directions will be semantically meaningful, and it is hard to know which direction controls a particular property of interest.}

% {However, the class-wise covariance matrix may be hard to estimate for medical data such as brain images, where differences between classes (e.g., AD and CN) are subtle.}

% data with some specific changes compared to the original data, as augmentations

\begin{figure}[t!]
    \centering
    \includegraphics[scale = 0.6]{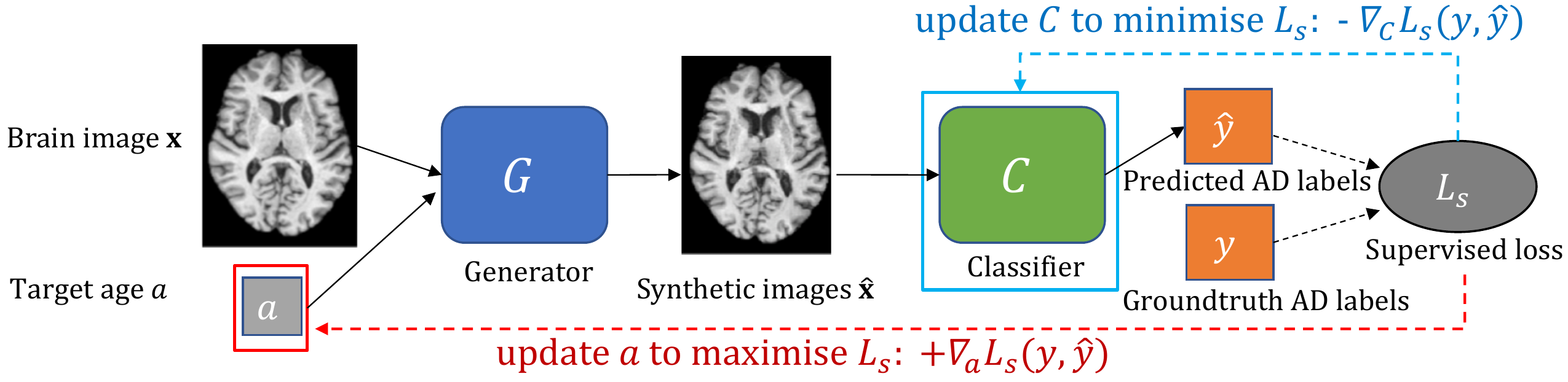}
    \caption{A schematic of the adversarial classification training. The pre-trained generator $G$ takes a brain image $\mathbf{x}$ and a target age $a$ as input and outputs a synthetically aged image $\hat{\mathbf{x}}$ that corresponds to the target age $a$. The classifier $C$ aims to predict AD label for a given brain image. To utilise  $G$ to improve  $C$, we formulate an adversarial game between  $a$ (in  \red{red} box) and $C$ (in \textcolor{cyan}{cyan} box), where $a$ and $C$ are updated alternatively and iteratively using $L_{1}$ and $L_{2}$, respectively (see Sec.~\ref{sec: Adversarial classification training with a pre-trained generator}). Note  $G$ is frozen.}
    % \tiantodo{I add arrows to show update. Does it illustrate the process more clear? I will modify details to make it look better later.}
    \label{fig: schematic of adversarial classification training}
\end{figure}

In this work, we consider the scenario that we have a classifier which we want to improve (e.g. an image-based classifier of Alzheimer's Disease (AD) given brain images). We are also given some data and a pre-trained generative model that is able to create new data given an image as input and conditioning factors that can alter corresponding attributes in the input. For example, the generative model can alter the brain age of the input. We propose an approach to guide a pre-trained generative model to generate the \textit{most effective} counterfactuals via an adversarial game between the input \textit{conditioning factor} of the generator and the downstream classifier, {where we use gradient backpropagation to update the \textit{conditioning factor} and the \textit{classifier} alternatively and iteratively.} 
%Inspired by adversarial attack~\citep{szegedy2013intriguing}, the idea is similar to a recent work~\citep{pervin2021adversarial}. However, \cite{szegedy2013intriguing,pervin2021adversarial} focused on  unperceivable perturbations, while we focus on finding hard generative factors.
A schematic of the proposed approach is shown in Fig.~\ref{fig: schematic of adversarial classification training}.

Specifically, we choose the classification of AD as the downstream task and utilise a pre-trained brain ageing synthesis model to improve the AD classifier.  {The brain ageing generative model used in this paper is adopted from a recent work~\citep{xia2021learning}, which takes a brain image and a target age as inputs and outputs an aged brain image\footnote{Code is available at https://github.com/xiat0616/BrainAgeing}}. We show that the proposed approach can improve the test accuracy  of the AD classifier. We also demonstrate that it can be used in a  {\textit{continual learning}\footnote{Deep models continuously learn based on input of new data while preserving previously learnt knowledge.} context to alleviate \textit{catastrophic forgetting}, i.e. deep models forget what they have learnt from previous data when training on new given data, and can be used to alleviate \textit{spurious correlations}, i.e. two variables appear to be causally related to one another but in fact they are not.} Our contributions can be summarised as follows:

\noindent \textbf{1.} We propose an approach to utilise a pre-trained generative model for a classifier via an adversarial game between \textit{conditional input}  and the \textit{classifier}. {To the best of our knowledge, this is the first approach that formulates such an adversarial scheme to utilise pre-trained generators in medical imaging.}
% \tiantodo{Should we say this?}
  
\noindent \textbf{2.} {We improve a recent brain ageing synthesis model by involving Fourier encoding to enable gradient backpropagation to update \textit{conditional factor} and demonstrate the effectiveness of our approach on the task of AD classification.}

\noindent \textbf{3.}  We consider the scenario of using generative models in a \textit{continual learning} context and show that our approach can help alleviate \textit{catastrophic forgetting}.

\noindent \textbf{4.} We apply the brain ageing synthesis model for brain \textit{rejuvenation} synthesis and demonstrate that the proposed approach has the potential to alleviate \textit{spurious correlations}.

\section{Methodology}
\label{sec: chap 6 methodology}
\subsection{Notations and problem overview}

% In the rest of this paper, we use $\mathbf{bold}$ notations for vectors/images. 
We denote an image as $\mathbf{x} \sim \mathcal{X}_{}$, and a conditional generative model $G$ that takes an image $\mathbf{x}$ and a conditional vector ${\mathbf{v}}$ as input and generates a counterfactual $\mathbf{\hat{x}}_{}$ that corresponds to $\mathbf{v}$: $\mathbf{\hat{x}}_{}=G(\mathbf{x}, \mathbf{v})$. For each $\mathbf{x}$, there is a label $y \sim \mathcal{Y}$. 
% In the rest of this paper, we use $\mathbf{bold}$ notations for vectors/images. We denote an image as $\mathbf{x}$ (and $\mathcal{X}$ their distribution such that $\mathbf{x_{a}}\sim \mathcal{X}_{a})$, where $a$ represents the corresponding \textit{attributes} of $\mathbf{x}_{a}$. For instance, if $\mathbf{x_{a}}$ is the face image of a person, then $a$ could be his/her age, gender, weight, etc. Then we define a conditional generative model \textit{G} that takes  an image $\mathbf{x}_{a}$ and a target  attribute vector $\mathbf{a}_{tar}$, and generates a counterfactual $\mathbf{\hat{x}}_{a_{tar}}$ that corresponds to target attributes $a_{tar}$: $\mathbf{\hat{x}}_{a_{tar}}=G(\mathbf{x}_{a}, \mathbf{a}_{tar})$. For each $\mathbf{x_{a}}$, there exits a label $y$ ($y \sim \mathcal{Y}$). 
% The goal is to learn a model $C_{\theta}$: $\mathcal{X_{a}} \rightarrow \mathbf{Y}$, parameterized by $\theta$, that minimizes the 
We define a  classifier $C$ that predicts the label $\hat{y}$ for given $\mathbf{x}_{}$: $\hat{y}=C(\mathbf{x}_{})$.
% Note here the label $y$ does not need to equal to the conditional factor $a$. For instance, if $\mathbf{x}$ is a face image of a person, then $v$ could be his/her age, and $y$ could be his/her gender. 
% Our solution is to find the counterfactuals that are the \textit{most helpful} to improve $C$. Inspired by the mini-max mechanism in GANs~\cite{goodfellow2014generative} and adversarial learning~\cite{goodfellow2014explaining,miyato2018virtual}, we formulate an adversarial game between the conditional attributes $a$ and the model $C$. That is, try to find the conditional attributes that result in \textit{hard} counterfactuals for $C$, and improve $C$ on these hard samples. This strategy can viewed as finding the `weakness' of the model $C$ and purposely forcing $C$ to overcome its `weakness'.
In this paper, $\mathbf{x}$ is a  brain image,  $y$ is the AD diagnosis of $\mathbf{x}$,  and $\mathbf{v}$ represents the target age $a$ and AD diagnosis on which the generator $G$ is conditioned.  {We select age and AD status to be conditioning factors as they are major contributors to brain ageing.} We use a brain ageing generative model as $G$, and a VGG\footnote{A popular deep learning neural network that has widely been used for classification.}-based~\citep{simonyan2015very} AD classification model as $C$. Note that we only change the target age $a$ in this paper, thus we write the generative process as $\hat{\mathbf{x}}=G(\mathbf{x}, a)$ for simplicity. 

Suppose a pre-trained  $G$ and a $C$ are given,  the question we want to answer is: \lq\lq \textit{How can we use  $G$ to improve $C$ in a (data) efficient manner}\rq\rq? To this end, {we propose an approach to utilise  $G$ to improve $C$ via an adversarial game with \textit{gradient backpropagation} to update $a$ and $C$ alternatively and iteratively.}

% \tiantodo{I changed the question into the description of the method}

% {Define a and V here.}

\subsection{Fourier encoding for conditional factors}
\label{sec: fourier eocnding for conditonal factors}

 {The proposed approach requires backpropagation of gradient to the conditional factor to find the hard counterfactuals.} However, the original brain ageing synthesis model~\citep{xia2021learning} used \textit{ordinal encoding} to encode the conditional age and AD diagnosis, where the encoded vectors are discrete in nature and need to maintain a certain shape, which hinders  gradient backpropagation to update these  vectors.

To enable gradient backpropagation to update the conditional vectors,  we propose to use \textit{Fourier encoding} \citep{tancik2020fourier,mildenhall2020nerf} to encode the conditional attributes, i.e.,~age and heath state (diagnosis of AD). The effectiveness of Fourier encoding has been experimentally shown in \cite{tancik2020fourier,mildenhall2020nerf}, and we also achieved similar synthetic results with Fourier encoding as the original model.  The key idea of Fourier encoding is to map low-dimensional vectors to a higher dimensional domain using a set of sinusoids. For instance, if we have a \textit{d}-dimensional vector which is normalised into $[0,1)$, $\mathbf{v} \in [0,1)^{d}$, then the encoded vector can be represented as~\cite{tancik2020fourier}:

\begin{equation}
    \mathbf{\gamma}(\mathbf{v}) = [p_{1}\cos(2\pi \mathbf{b_{1}^T \mathbf{v})}, p_{1}\sin(2\pi \mathbf{b_{1}^T} \mathbf{v}), ..., p_{m}\cos(2\pi \mathbf{b_{m}^T} \mathbf{v}), p_{m}\sin(2\pi \mathbf{b_{m}^T}\mathbf{v})],
\end{equation}
where $\mathbf{b_{j}}$ can be viewed as the Fourier basis frequencies, and $p^2_{j}$ the Fourier series coefficients.

In this work, the vector $\mathbf{v}$ represents the target age $a$ and the health status (AD diagnosis), and $d=2$.  In our experiments, we set $p^2_{j}=1$ for $j=1, ..., m$, and $\mathbf{b_{j}}$ are independently and randomly sampled from a Gaussian distribution, $\mathbf{b_{j}} \sim \mathcal{N}(\mu_{scale}*\mathbf{I}, 0)$, where $\mu_{scale}$ is  set to 10. We set $m=100$ and the resulting $\gamma(\mathbf{v})$ is 200-dimensional. After encoding, the generator $G$ takes the encoded vector $\gamma(\mathbf{v})$  as input.

The use of Fourier encoding offers two advantages. First,  \cite{xia2021learning} encoded age and health state into two vectors and had to use two MLPs to embed the encoded vectors into the model. This may not be a big issue when the number of factors is small. However,  {extending the generative model to be conditioned on tens or hundreds of factors will increase the memory and computation costs significantly}. With Fourier encoding, we can encode all possible factors into a single vector, which offers more flexibility to scale the model to multiple conditional factors. Second, Fourier encoding allows us to compute the gradients with respect to the input vector $\mathbf{v}$ or certain elements of $\mathbf{v}$, since the encoding process is differentiable. As such, we replace the \textit{ordinal encoding} with \textit{Fourier encoding} for all experiments. The generative model $G$ takes $\mathbf{v}$ as input: $\hat{\mathbf{x}}=G(\mathbf{x}, \mathbf{v})$, where $\mathbf{v}$ represents target age and health state. Since we only change the target age $a$ in this paper, we write the generative process as $\hat{\mathbf{x}}=G(\mathbf{x}, a)$ for simplicity.

\subsection{Adversarial counterfactual augmentation}
\label{sec: Adversarial classification training with a pre-trained generator}

Suppose we have a conditional generative model $G$ and a classification model $C$. The goal is to utilise $G$ to improve the performance of $C$. To this end, we propose an approach consisting of three steps: \textit{pre-training}, \textit{hard sample selection} and \textit{adversarial classification training}. {A schematic of the adversarial classification training is presented in Fig.~\ref{fig: schematic of adversarial classification training}.} Algorithm~\ref{alg: the adversarial algorithm} summarises the steps of the method. Below we describe each step in detail.

\noindent \textbf{Pre-training.} 
% We first pre-train the classification model $C$ on the training dataset $D_{train}: \{\mathcal{X}_{train}, \mathcal{Y}_{train}\}$. Specifically, 
The generative model is pre-trained using the same losses as in \cite{xia2021learning} except that we use Fourier encoding to encode age and AD diagnosis. Consequently, we obtain a pre-trained $G$ that can generate counterfactuals conditioned on given target ages $a$: $\mathbf{\hat{x}} = G(\mathbf{x}, {a})$.

The classification model $C$ is a VGG-based network~\citep{simonyan2015very} trained to predict the AD diagnosis from brain images, optimised by minimising:
\begin{equation}
\label{eq: pre-train classifier}
    L_{pre-train} = \mathbb{E}_{\mathbf{x}\sim \mathcal{X}_{train}, \mathbf{y} \sim \mathcal{Y}_{train}} L_{s}(C(\mathbf{x}), y ), 
\end{equation}
where $L_{s}(.)$ is a supervised loss (binary cross-entropy loss in this paper),  $\mathbf{x}$ is a brain image, and $y$ is its ground-truth AD label. To note that if the pre-trained $G$ and $C$ are available in practice, we could avoid the pre-training step.

\noindent \textbf{Hard sample selection.} \cite{liu2021just,feldman2020neural} suggested that training data samples have different \textit{influence} on the training of a supervised model, i.e.,~some training data are \textit{harder} for the task and are more \textit{effective} to train the model than others.  \cite{liu2021just} propose to up-sample, \textit{i.e.} duplicate, the \textit{hard} samples as a way to improve the model performance.
Based on these observations, we propose a similar strategy to \cite{liu2021just} to select these \textit{hard} samples: we record the classification errors of all training samples for the pre-trained $C$ and then select $N=100$ samples with the highest errors. The selected  \textit{hard} samples are denoted as $D_{hard}$: $\{X_{hard}, Y_{hard}\}$.

\noindent \textbf{Adversarial classification training.}
\citep{bowles2018gan,frid2018gan,dar2019image} augmented datasets by randomly generating a number of synthetic data with pre-trained generators. Similar to training samples, some synthetic data could be more \textit{effective} for downstream tasks than others. Here we assume that if a synthetic data sample is \textit{hard}, then it is more \textit{effective} for training. We propose an adversarial game to find the \textit{hard} synthetic data to boost $C$. 

\begin{algorithm}[t!]
\caption{Adversarial counterfactual augmentation with a pre-trained $G$.} \label{alg: the adversarial algorithm}
\begin{algorithmic}
\State {\textbf{Input:} Training set $D_{train}$; hyperparameter $k$, $N$; a pre-trained $G$;  $C$.}
% \Require 
\State \textbf{Pre-training:} 
\State \indent  {1. Train the classifier $C$ on $D_{train}$ (Eq.~\ref{eq: pre-train classifier}). }
\State \textbf{Hard sample selection: }
\State \indent {2. Select $N$  samples from $D_{train}$ that result in highest classification errors} 
\State  \indent {for $C$, denoted as $D_{hard}$.}
\State \textbf{Adversarial classification training: }
\State \indent  {3. Randomly initialize target ages $a$, and obtain initial synthetic data. }
\State \indent \textbf{For $k$ do}
\State \indent  \indent {4. Update $a$ in the direction to maximise classification error (Eq.~\ref{eq: loss of update attribute}).}
\State \indent \indent {5. Obtain synthetic images with $D_{hard}$ and the updated $a$, denoted as $D_{syn}$.}
\State  \indent \indent {6. Update $C$ to optimise Eq.~\ref{eq: update classifier} on $D_{train} \cup D_{syn}$ for one epoch.}
\end{algorithmic}
\end{algorithm}

Specifically, let us first define the classification loss for  synthetic data as:
\begin{equation}
 L_{C} = \mathbb{E}_{\mathbf{x}\sim X_{hard}, {y}\sim Y_{hard}}L_{s}(C(\mathbf{\hat{x}}), y),
    % L_{C} = L_{s}(C(\mathbf{\hat{x}}, y)),
\end{equation}
where  $\mathbf{\hat{x}}$ is a generated sample  conditioned on the target age $a$: $\mathbf{\hat{x}}=G(\mathbf{x}, a)$, and $y$ is the ground-truth AD label for $\mathbf{{x}}$. Here we assume that  changing target age does not change the AD status, thus $\mathbf{{x}}$ and $\mathbf{\hat{x}}$ have the same AD label.

Since the encoding of age $a$ is differentiable (see Section~\ref{sec: fourier eocnding for conditonal factors}), we can obtain the gradients of  $L_{C}$ with respect to $a$ as: $\nabla_{a}L_{C} = \nabla_{a}[L_{s}(C(G(\mathbf{x}, a)), y)]$,
and update $a$ in the direction of \textit{maximising} $L_{C}$ by: $\tilde{a} = a+\gamma_{a} \nabla_{a}L_{C}$, where $\gamma_{a}$ is the step size (learning rate) for updating $a$. Formally, the optimization function of $a$ can be written as:
\begin{equation}
\label{eq: loss of update attribute}
    L_{1} = \max_{a} \E_{\mathbf{x}\sim X_{hard}, {y}\sim Y_{hard}}L_{s}(C(\mathbf{\hat{x}}), y).
\end{equation}
Then we could obtain a set of synthetic data using the updated $\tilde{a}$: $\hat{\mathbf{x}}_{syn}=G(\mathbf{x}_{},\tilde{a})$ where $\mathbf{x}\sim X_{hard}$, denoted as $D_{syn}: \{X_{syn}, Y_{syn}\}$.

The classifier $C$ is updated by optimising:
\begin{equation}
\label{eq: update classifier}
    L_{2} = \min_{C}\E_{\mathbf{x}\sim X_{combined}, y\sim Y_{combined} } L_{s} (C(\mathbf{x}), y),
\end{equation}
where $D_{combined}$: $\{ X_{combined}, Y_{combined}\}$ is a combined dataset consisting of the training dataset and synthetic dataset: $\{X_{combined}, Y_{combined}\} = \{X_{train}\cup X_{syn}, Y_{train}\cup Y_{syn} \}$. Similar to \cite{liu2021just}, we update $C$ on $D_{combined}$ instead of $D_{syn}$ as we found updating $C$ only on $D_{syn}$ can cause {catastrophic forgetting}~\citep{kirkpatrick2017overcoming}.

The adversarial game is formulated by alternatively and iteratively updating $a$ and classifier $C$ via Eqs.~\ref{eq: loss of update attribute} and \ref{eq: update classifier}, respectively.  In practice, to prevent $a$ from going to unsuitable ages, we clip it to be in [60, 90] after every update.

\noindent \textbf{Updating $a$ vs. updating $G$.} Note here the adversarial game is formulated between $a$ and $C$, instead of $G$ and $C$. This is because training $G$ against $C$ allows $G$ to change its latent space without considering image quality, and the output of $G$ could be unrealistic. Please refer to Section~\ref{sec: train g vs c} for more details and results.

\noindent  {\textbf{Counterfactual augmentation vs. conventional augmentation.} Here we choose to augment data counterfactually instead of applying conventional augmentation techniques. This is because that the training and testing data are already pre-processed and registered to MNI 152, and in this case conventional augmentations do not introduce helpful variations. Please refer to Section~\ref{sec: Counterfactual augmentation v.s. conventional augmentation} for more details and results.}

\subsection{Adversarial classification training in a \textit{continual learning} context}
% \label{sec: links to continual learrning}
Most previous works~\citep{bowles2018gan,antoniou2017data,frid2018gan,frid2018synthetic,shin2018medical,dar2019image} that used pre-trained deep generative models for augmentation focused on generating a large number of synthetic samples, and then merged the synthetic data with the original dataset and trained the downstream task model (\textit{e.g.} a classifier) on this augmented dataset. However, this requires training the task model from scratch, which could be inconvenient. For instance, if we suddenly decided to generate some new synthetic data for augmentation, we would have to retrain the task model from scratch. Furthermore, if the size of the original dataset is large, then the number of synthetic samples can be huge, which would make the training process extremely expensive and time-consuming. Thus, in practice, we need to consider cases where we aim to improve a pre-trained classifier with synthetic data but without retraining the whole model from scratch. We design the proposed procedure in such a way that allows us to use the pre-trained $G$ to improve $C$ flexibly.

% \red{Not sure if should remove this paragraph}

In Section~\ref{sec: Adversarial classification training with a pre-trained generator}, after we obtain the synthetic set $D_{syn}$, we choose to update the classifier $C$ on the augmented dataset $D_{syn} \cup D_{train}$, instead of $D_{syn}$ (stage 6 in Algorithm~\ref{alg: the adversarial algorithm}). This is because re-training the classifier only on the $D_{syn}$ would result in \textit{catastrophic forgetting}~\citep{kirkpatrick2017overcoming}, \textit{i.e.}~a phenomenon where deep neural networks tends to forget what it has learnt from previous data when being trained on new data samples. To alleviate catastrophic forgetting, efforts have been devoted to developing approaches to allow artificial neural networks to learn in a sequential manner~\citep{delange2021continual,parisi2019continual}. These approaches are known as \textit{continual learning}~\citep{delange2021continual,DBLP:journals/corr/abs-1902-10486,lopez2017gradient}, \textit{lifelong learning}~\citep{chen2018lifelong,aljundi2017expert}, \textit{sequential learning}~\citep{mccloskey1989catastrophic,aljundi2018selfless}, or \textit{incremental learning}~\citep{chaudhry2018riemannian,gepperth2016bio}. Despite different names and focuses, the main purpose of these approaches is to overcome catastrophic forgetting and to learn in a sequential manner. 

\begin{algorithm}[!tb]
\caption{Adversarial classification learning with $D_{store}$.} \label{alg: effect of M and N}
\begin{algorithmic}
\State \textbf{ Input:} Training dataset $D_{train}$; hyperparameter $M$, $N$, $k$; a pre-trained generator $G$; a pre-trained classifier model $C$.
% \Require 
\State \textbf{Construct $D_{store}$: }
\State \indent {1. Randomly select $M\%$ data from $D_{train}$, denoted as $D_{store}$.}
\State \textbf{Hard sample selection}
\State \indent {2. Select $N$ samples from $D_{store}$ that result in highest classification errors for $C$, denoted as $D_{hard}$.}
\State \textbf{Adversarial training: }
\State \indent  {3. Randomly initialise target ages $a$, and obtain initial synthetic data. }
\State \indent \textbf{For $k$ do}
\State \indent  \indent {4. Update $a$ in the direction to maximise classification error (Equation~\ref{eq: loss of update attribute}).}
\State \indent  \indent {5. Obtain synthetic images with $D_{hard}$ and the updated $a$, denoted as $D_{syn}$.}
\State \indent  \indent {6. Update $C$ to minimise the classification error on $D_{store} \cup D_{syn}$ (Equation~\ref{eq: update classifier}).}
\end{algorithmic}
\end{algorithm}

If we consider the generated data as new samples, then the update of the pre-trained classifier $C$ can be viewed as a \textit{continual learning} problem, \textit{i.e.} how to learn \textit{new} knowledge from the synthetic set $D_{syn}$ without forgetting \textit{old} knowledge that is learnt from the original training data $D_{train}$. To alleviate  catastrophic forgetting, we re-train the classifier on both the synthetic dataset $D_{syn}$ and the original training dataset $D_{train}$. This strategy is known as \textit{memory replay} in continual learning~\citep{robins1995catastrophic,van2020brain} and was also used in other augmentation works~\citep{liu2021just}. The key idea is to store previous data in a \textit{memory buffer} and \textit{replay} the saved data to the model when training on new data. However, it could be expensive to store and revisit all the training data, especially when the data size is large~\citep{van2020brain}. In Section~\ref{sec: Adversarial classification training in continual learning context}, we perform experiments where we only provide a portion ($M\%$) of training data to the classifier when re-training with synthetic data (to simulate the \textit{memory buffer}). We want to see whether \textit{catastrophic forgetting} would happen or not when only a portion ($M\%$) of training data is provided, and if so, how much it affects the test accuracies. Algorithm~\ref{alg: effect of M and N} summarises the steps of the method in the \textit{continual learning} context.

\section{Experimental setup}
\label{sec: chap 6 experimental setup}
\noindent \textbf{Data.} {We use the  ADNI dataset~\citep{petersen2010alzheimer} for experiments. We select 380 AD and 380 CN (control normal) T1 volumes between 60 and 90 years old. 
We split the AD and CN data into training/validation/testing sets with 260/40/260 volumes, respectively.  All volumetric data are skull-stripped using DeepBrain\footnote{https://github.com/iitzco/deepbrain}, and linearly registered to MNI 152 space using FSL-FLIRT~\citep{woolrich2009bayesian}. We normalise brain volumes by clipping the intensities to $[0,V_{99.5}]$, where $V_{99.5}$ is the $99.5\%$ largest intensity value within each volume, and then rescale the resulting intensities to the range $[-1,+1]$. We select the middle 60 axial slices from each volume and crop each slice to the size of $[208,160]$, resulting in 31200 training, 4800 validation and 9600 testing slices.}

\noindent \textbf{Implementation.} 
The generator is trained the same way as in \cite{xia2021learning}, except we replace \textit{ordinal encoding} with \textit{Fourier encoding}. We pre-train the classifier for 100 epochs. The experiments are performed using Keras and Tensorflow. We train pre-trained classifiers $C$ with Adam with a learning rate of 0.00001 and decay of 0.0001. During adversarial learning, the step size of $a$ is tuned to be 0.01, and the learning rate for $C$ is 0.00001. The experiments are performed using a NVIDIA Titan X GPU.

\noindent \textbf{Comparison methods.} We compare with the following baselines:

\noindent \textbf{1.~\textit{Na\"ive}}: We directly use the pre-trained $C$ for comparison as the \textit{lower bound}.

\noindent \textbf{2.~\textit{RSRS}}:  Random Selection + Random Synthesis. We randomly select $N=100$ samples from the training set $D_{train}$, denoted as $D_{rand}$, and then use the generator $G$ to randomly generate $N_{synthesis}=5$ synthetic samples for each sample in $D_{rand}$, denoted as $D_{syn}$. Then we train the classifier on the combined dataset $D_{train}\cup D_{syn}$ for $k=5$ steps. This is the typical strategy used by most previous works~\citep{bowles2018gan,frid2018gan,dar2019image}.
% We randomly select N training samples as $D_{rand}$ and randomly generate counterfactuals from $D_{rand}$.

\noindent \textbf{3.~\textit{HSRS}}: Hard Selection + Random Synthesis. We select $N=100$ hard samples from $D_{train}$ based on their classification errors of $C$, denoted as $D_{hard}$, and then use the generator $G$ to randomly generate $N_{synthesis}=5$ synthetic samples for each sample in $D_{hard}$, denoted as $D_{syn}$. Then we train the classifier on the combined dataset $D_{train}\cup D_{syn}$ for $k=5$ steps.

\noindent \textbf{4.~\textit{RSAT}}:  Random Selection + Adversarial Training. We randomly select $N=100$ samples from the training set $D_{train}$, denoted as $D_{rand}$, and then use the adversarial training strategy to update the classifier $C$, as described in Sec.~\ref{sec: Adversarial classification training with a pre-trained generator}.. The difference between RSAT and our approach is that we select hard samples for generating counterfactuals, while RSAT uses random samples.

\noindent \textbf{5.~\textit{JTT}}: Just Train Twice ~\citep{liu2021just} record samples that are misclassified by the pre-trained classifier, obtaining an error set. Then they construct an oversampled dataset $D_{up}$ that contain examples in the error set $\lambda_{up}$ times and all other examples once. Finally, they train the classifier on the oversampled dataset $D_{up}$. In this paper, we set $\lambda_{up}=2$ as we found large $\lambda_{up}$ results in bad performance. We also found the original learning rate (0.01) used for the second training stage results in very bad performance and set it to be 0.00001.

%There are 5 heading levels

\section{Results and Discussion}
\label{sec: chap6 results and discussion}

\subsection{Improving the performance of classifiers. }
\label{sec: main results for adversarial training}

\subsubsection{Comparison with baselines.}
We first compare our method with baseline approaches by evaluating the test accuracy of the classifiers. We set $N=100$ and $k=5$ in experiments.  We pre-train $C$ for 100 epochs and $G$ as described in Section~\ref{sec: chap 6 experimental setup}. The weights of the pre-trained  $C$ and the pre-trained $G$  are the same for all methods. For a fair comparison, the total number of synthetically generated samples is fixed to 500 for \textit{RSRS}, \textit{HSRS}, \textit{RSAT} and our approach. For \textit{JTT}, there are 2184 samples mis-classified by $C$ and oversampled. We initialize $a$ randomly between real ages of original brain images $\mathbf{x}$ and maximal age (90 yrs old). 

\begin{table}[tb]
\centering
\caption{Average test accuracies of models trained via our procedure and baselines. We first present the average test accuracies for different age groups with AD (column 2-4) or CN (column 5-7) and then present the average test accuracies for the whole testing set (column 8). For each method, the \textit{worst-group} performance  is shown in \textit{italic}. For each age group, \textit{i.e.}~each column, the \textbf{best} performance is shown in \textbf{bold}. We also report the number of testing images for each age group.  }
\begin{tabular}{c|ccc|ccc|c}
\hline
Acc \%       & \multicolumn{3}{c|}{CN}                                & \multicolumn{3}{c|}{AD}         & All           \\ \hline
Age group      & 60-70yrs         & 70-80yrs         & 80-90yrs                  & 60-70yrs         & 70-80yrs         & 80-90yrs         & overall       \\
Test group size & 1540          & 1600          & 1660                   & 1720          & 1540          & 1540          & 9600          \\ \hline
Na\"ive        & 85.2          & 91.5          & \textit{70.7}          & 92.5        & 94.2         & \textbf{97.1} & 88.4          \\ \hline
RSRS         & 86.0          & 90.4          & \textit{73.8}          & 87.3          & 95.1          & 90.0          & 87.0          \\ \hline
HSRS         & 85.6         & 91.1          & \textit{80.4}          & 89.8           & 93.8          & 96.9          & 89.5          \\ \hline
RSAT         & 86.1          & 93.1          & \textit{81.5}          & 91.8          & 96.0          & 95.7          & 90.6          \\ \hline
JTT          & 83.9          & \textbf{94.2} & \textit{80.1}          & \textbf{92.8} & 90.8          & 93.7        & 89.2          \\ \hline
Proposed     & \textbf{86.4} & 93.7          & \textit{\textbf{83.4}} & 91.5          & \textbf{96.5} & 95.7       & \textbf{91.1} \\ \hline
\end{tabular}
\label{tab: accuracy for main experiment}
\end{table}

From Table~\ref{tab: accuracy for main experiment} we can observe that our proposed procedure achieves the best overall test accuracy, followed by baseline \textit{RSAT}. This demonstrates the advantage of adversarial training between the conditional factor (target age) $a$ and the classifier. On top of that, it shows that selecting \textit{hard} examples for creating augmented synthetic results helps, which is also demonstrated by the improvement of performance of \textit{HSRS} over \textit{Na\"ive}. We also observe that \textit{JTT}~\citep{liu2021just} improves the classifier performance over \textit{Na\"ive}, showing the benefit of up-sampling \textit{hard} samples.  In contrast, baseline \textit{RSRS} achieves the lowest overall test accuracy, even lower than that of \textit{Na\"ive}. This shows that randomly synthesising counterfactuals from randomly selected samples could result in synthetic images that are harmful to the classifier.

Furthermore, we observe that for all methods, the \textit{worst-group} performances are achieved on the 80-90 CN group. A potential reason could be: as age increases, the brains shrink, and it is harder to tell if the ageing pattern is due to AD or caused by normal ageing. Nevertheless, we observe that for this \textit{worst group}, our proposed method still achieves the best performance, followed by $RSAT$. This shows that adversarial training can be helpful to improve the performance of the classifier, especially for \textit{hard} groups. The next best results are achieved by \textit{HSRS} and \textit{JTT}, which shows that finding hard samples and up-sampling or augmenting them was helpful to improve the \textit{worst-group} performance. We also observe the improvement of \textit{worst-group} performance for \textit{RSRS} over \textit{Na\"ive}, but the improvement is small compared to other baselines.

% We also measure the Area Under Curve (AUC) values for all methods, as presented in Table~\ref{tab: AUC results}. We can observe that our approach achieves the highest overall AUC results. 

We also report the \textit{precision} and \textit{recall} for all methods, as presented in Table~\ref{tab: precision and recall results}. We can observe that our approach achieves the highest overall precision and recall rersults.

% \begin{table}[tb]
% \centering
% \caption{The test Area Under the ROC Curve (AUC)~\citep{bradley1997use} values for all methods. We first present the AUC for different age groups (column 2-4), and then present the AUC for all testing data (column 5). For each group, the \textbf{best} results are shown in \textbf{bold}.}
% \begin{tabular}{c|ccc|c}
% \hline
% AUC      & 60-70yrs          & 70-80yrs          & 80-90yrs          & Overall        \\ \hline
% Na\"ive    & 0.954          & 0.968          & 0.903          & 0.931          \\ \hline
% RSRS     & 0.932          & 0.977          & 0.904          & 0.928          \\ \hline
% HSRS     & 0.958          & 0.975          & \textbf{0.921} & 0.954          \\ \hline
% RSAT     & 0.955          & 0.981          & 0.912          & 0.957          \\ \hline
% JTT      & 0.957          & 0.978          & 0.914          & 0.952          \\ \hline
% Proposed & \textbf{0.961} & \textbf{0.988} & 0.917          & \textbf{0.960} \\ \hline
% \end{tabular}
% \label{tab: AUC results}
% \end{table}

\begin{table}[tb]
\centering
\caption{The test precision and recall values for all methods. We first present the precision for different age groups (column 2-4) and  all testing data (column 5), and then present the recall for different age groups (column 6-8) and all testing data (column 9). For each group, the \textbf{best} results are shown in \textbf{bold}. }
\begin{tabular}{c|cccc|cccc}
\hline
Age Range & 60-70          & 70-80          & 80-90          & Overall        & 60-70          & 70-80          & 80-90          & Overall        \\ \hline
Metrics   & \multicolumn{4}{c|}{Precision}                                    & \multicolumn{4}{c}{Recall}                                        \\ \hline
Naive     & 0.875          & 0.914          & 0.761          & 0.842          & 0.925          & 0.942          & \textbf{0.971} & \textbf{0.945} \\ \hline
RSRS      & 0.874          & 0.905          & 0.768          & 0.844          & 0.873          & 0.951          & 0.900          & 0.906          \\ \hline
HSRS      & 0.874          & 0.910          & 0.826          & 0.866          & 0.898          & 0.938          & 0.969          & 0.933          \\ \hline
RSAT      & 0.881          & 0.930          & 0.832          & 0.877          & 0.918          & 0.960          & 0.957          & 0.943          \\ \hline
JTT       & 0.865          & \textbf{0.938} & 0.822          & 0.868          & \textbf{0.928} & 0.908          & 0.960          & 0.924          \\ \hline
Proposed  & \textbf{0.883} & 0.936          & \textbf{0.848} & \textbf{0.885} & 0.915          & \textbf{0.965} & 0.965          & \textbf{0.945} \\ \hline
\end{tabular}
\label{tab: precision and recall results}
\end{table}

In summary, the quantitative results show that it is helpful to find and utilise \textit{hard} counterfactuals for improving the classifier.

\subsubsection{Train \textit{G} against \textit{C}}
\label{sec: train g vs c}
We choose to formulate an adversarial game between the conditional generative factor $a$ (the target age) and the classifier $C$, instead of between the generator $G$ and the classifier $C$. This is because we are concerned that an adversarial game between $G$ and $C$ could result in unrealistic outputs of $G$. In this section, we perform an experiment to investigate this. 

Specifically, we define an optimization function:
\begin{equation}
\label{eq: train G against C}
    L_{G} = \max_{G} \E_{\mathbf{x}\sim X_{train}, \mathbf{y}\sim Y_{train}} L_{s}(C(G(\mathbf{x},a )), y),
\end{equation}
where we aim to train $G$ in the direction of maximising the loss of the classifier $C$ on the synthetic data $G(\mathbf{x}, a)$. 

After every update of $G$, we construct a synthetic set $D_{syn}$ by generating {100} synthetic images from $D_{train}$, and update $C$ on $D_{train}\cup D_{syn}$ via Equation~\ref{eq: update classifier}. The adversarial game $G$  vs. $C$ is formulated by alternatively optimising Equation~\ref{eq: train G against C} and \ref{eq: update classifier} for {10} epochs.

\begin{figure}[tb]
    \centering
    \includegraphics[scale=0.46]{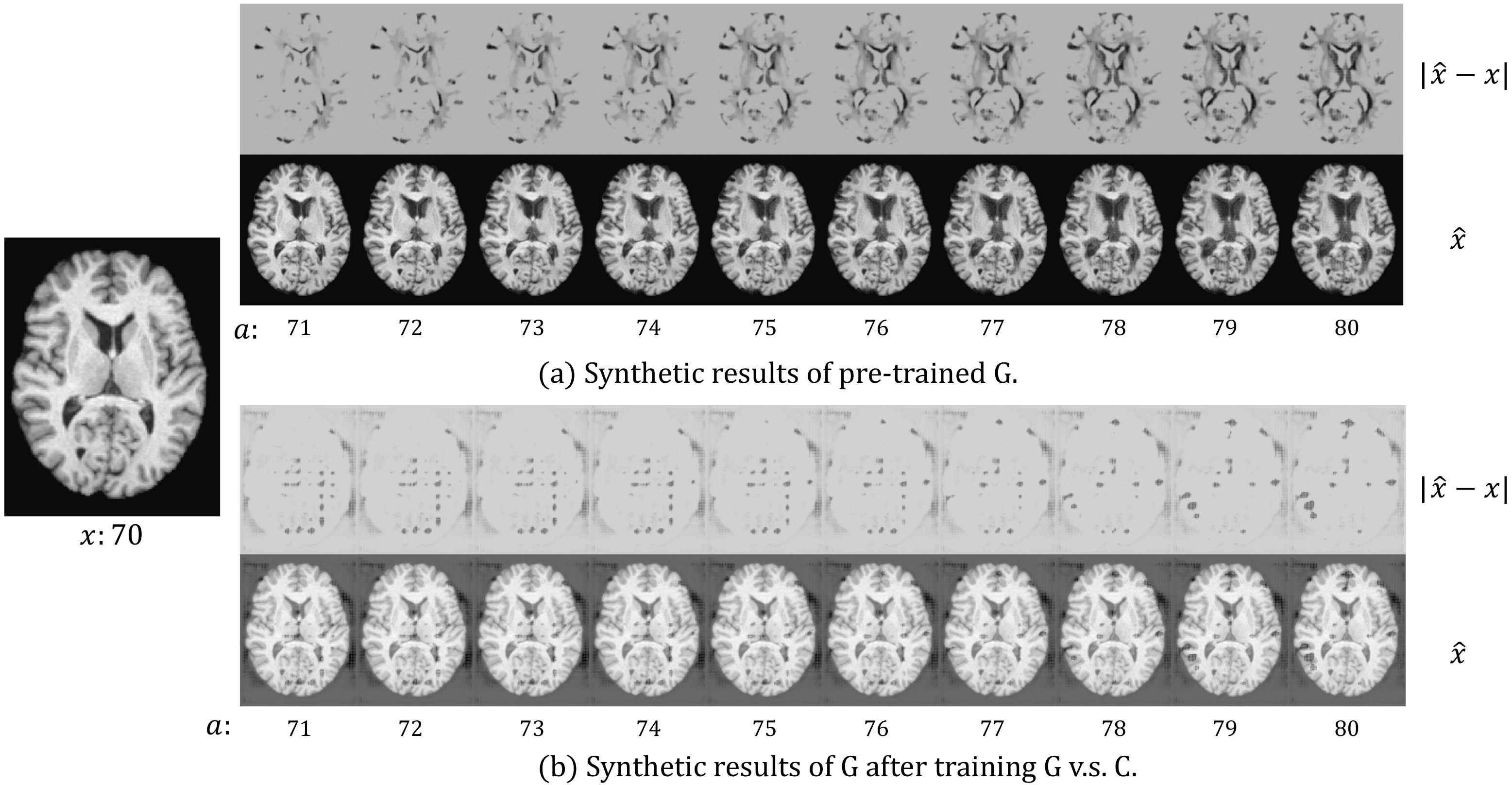}
    \caption{The synthetic results for a healthy (CN) subject $x$ at age 70:  (a) the results of the pre-trained $G$, \textit{i.e.}~before we train $G$ against $C$; (b) the results of $G$ after we train $G$ against $C$. We synthesise aged images $\hat{x}$ at different target ages $a$. We also {visualise} the difference between $x$ and $\hat{x}$, $|\hat{x}-x|$. For more details see text.}
    \label{fig: synthetic results after training G against C}
\end{figure}

In Figure~\ref{fig: synthetic results after training G against C}, we present the synthetic brain ageing progression of a CN subject before and after the adversarial training of $G$ vs. $C$. We can observe that after the adversarial training, the generator $G$ produces unrealistic results. This could be because there is no loss or constraint to prevent the generator $G$ from producing low-quality results. The adversarial game only requires the generator $G$ to produce images that are hard for the classifier $C$, and naturally, images of low quality would be hard for $C$. A potential solution could be to involve a GAN loss with a discriminator to improve the output quality, but this would make the training much more complex and require more memory and computations. We also measure the test accuracy of the classifier $C$ after training $G$ against $C$ to be $81.6\%$, which is much lower than the \textit{Na\"ive} method ($88.4\%$) and our approach ($91.1\%$) in Table~\ref{tab: accuracy for main experiment}. The potential reason is that $C$ is misled by the unrealistic samples generated by $G$.

\subsubsection{Effect of conventional augmentations for registered brain MRI data.}
\label{sec: Counterfactual augmentation v.s. conventional augmentation}
In this section, we test the effect of applying conventional augmentations, e.g. rotation, shift, scale and flip, to the training of the AD classifier. These are typical conventional augmentation techniques applied to computer vision classification task. Specifically, we train the classifier  the same way as \textit{Na\"ive}, except we augment training data with conventional augmentations.

Interestingly, we find that after applying rotation (range 10 degrees), shift (range 0.2), scale (range 0.2), and flip to augment the training data, the accuracy of the trained classifier drops from $88.4\%$ to $71.6\%$. We then measure accuracies when trained with each augmentation to be $74.1\%$ (rotation), $87.1\%$ (shift), $82.9\%$ (scale), and $87.8\%$ (flip). This could be because both training and testing data are already pre-processed, including registered to MNI 152, and these conventional augmentations do not introduce helpful variations to the training data but distract the classifier from focusing on subtle differences between AD and CN brains.

We also tried to train the classifier with MaxUp~\citep{gong2021maxup} with conventional augmentations. The idea of MaxUp is to generate a small batch of augmented samples for each training sample and train the classifier on the \textit{worst-performance} augmented sample. The overall test accuracy is $57.7\%$. This could be because that MaxUp tends to select the augmentations that distract the classifier from focusing on subtle AD features the most.
 
The results with conventional aumentations (+MaxUp) suggest that for the task of AD classification, when training and testing data are pre-processed well, conventional data augmentation techniques seem to not help improve the classification performance. Instead, these augmentations distract the classifier from identifying subtle changes between CN and AD brains. By contrast, the proposed procedure augment data in terms of semantic information, which could alleviate data imbalance and improve classification performance.

\subsection{Adversarial counterfactual augmentation in a \textit{continual learning} context}
\label{sec: Adversarial classification training in continual learning context}

\subsubsection{Results when re-training with a portion ($M\%$) of training data}
Suppose we have a pre-trained classifier $C$ and a pre-trained generator $G$, and we want to improve $C$ by using $G$ for data augmentation. However, after pre-training, we only store $M\%$ ($M\in (0,100]$) of the training dataset, denoted as $D_{store}$. During the adversarial training, we synthesise $N$ samples using the generator $G$, denoted as $D_{syn}$. Then we update the classifier $C$ on $D_{store}\cup D_{syn}$, using Equation~\ref{eq: update classifier} where $D_{combined}=D_{store}\cup D_{syn}$. The target ages are initialised and updated the same way as in Section~\ref{sec: main results for adversarial training}. Algorithm~\ref{alg: effect of M and N} illustrates the procedure in this section.

Table~\ref{tab: results when M changes} presents the test accuracies of our approach and baselines when $M$ changes. For \textit{Na\"ive-100}, the results are then same as in Table~\ref{tab: accuracy for main experiment}. For JTT, the original paper~\cite{liu2021just} retrained the classifier using the whole training set. Here we first randomly select $M \%$ training samples as $D_{store}$ and find misclassified data $D_{mis}$ within $D_{store}$ to up-sample, then we retrain the classifier on the augmented set. We can observe that when $M$ decreases, \textit{catastrophic forgetting} happens for all approaches. However, our method suffers the least from catastrophic forgetting, especially when $M$ is small. With $M=20 \%$ of training data for retraining, our approach achieves better results than \textit{Na\"ive}. This might be because the adversarial training between $a$ and $C$ tries to detect what is missing in $D_{store}$ and tries to recover the missing data by updating $a$ towards those directions. We observe that \textit{RSAT} achieves the second best results, only slightly worse than the proposed approach. Moreover, \textit{HSRS} and \textit{JTT} are more affected by catastrophic forgetting and achieve worse results. This might be because the importance of selecting \textit{hard} samples declines as $M$ decreases, since the $D_{store}$ becomes smaller.  

\begin{table}[!t]
\centering
\caption{Test accuracies of our approach and baselines when the ratio of the size $D_{store}$ vs. the size of $D_{train}$ changes. We can observe the decreases of test accuracies when $M$ decreases, which was due to the effect of \textit{catastrophic forgetting}. }
\begin{tabular}{c|ccccc}
\hline
Acc \%   & \multicolumn{5}{c}{$M\%$ } \\ \hline
Methods  & 1             & 10            & 20           & 50           & 100          \\ \hline
Na\"ive    & N/A           & N/A           & N/A          & N/A          & 88.4         \\ \hline
HSRS     & 75.6          & 81.4          & 84.5         & 87.4         & 89.5         \\ \hline
RSAT     & 84.2          & 85.8          & 87.2         & 88.6         & 90.6         \\ \hline
JTT      & 77.3           & 82.3         & 85.1       & 88.1          & 89.2         \\ \hline
Proposed & 84.8          & 86.8          & 88.5         & 89.4         & 91.1         \\ \hline
\end{tabular}
\label{tab: results when M changes}
\end{table}

These results demonstrate that our approach could alleviate \textit{catastrophic forgetting}. This could be helpful in cases where we want to utilise generative models to improve pre-trained classifiers (or other task models) without \textit{revisiting} all the training data (a \textit{continual learning} context). 

\subsubsection{Results when number of samples used for synthesis ($N$) changes}
We also performed experiments where we changed $N$, \textit{i.e.}~the number of samples used for generating counterfactuals. Specifically, we set $M=1$, \textit{i.e.}~only $1\%$ of original training data are used for re-training $C$, to see how many synthetic samples are needed to maintain good accuracy, especially when there are only a few training data stored in $D_{store}$. This is to see how \textit{efficient} the synthetic samples are in terms of training $C$ and alleviating \textit{catastrophic forgetting}. The results are presented in Table~\ref{tab: results when N changes}.

\begin{table}[!t]
\centering
\caption{Test accuracies when $N$ changes ($M=1$) of our approach and baselines.}
\begin{tabular}{c|cccc}
\hline
acc \%   & \multicolumn{4}{c}{\textit{N}}     \\ \hline
Methods  & 1    & 10   & 50   & 100  \\ \hline
HSRS     & 65.4 & 71.0 & 73.4 & 75.6 \\ \hline
RSAT     & 81.3 & 82.1 & 83.2 & 84.2 \\ \hline
Proposed & 82.1 & 82.9 & 84.1 & 84.6 \\ \hline
\end{tabular}
\label{tab: results when N changes}
\end{table}

From Table~\ref{tab: results when N changes}, we can observe that the best results are achieved by our method, followed by \textit{RSAT}. Even with only one sample for synthesis, our method could still achieve a test accuracy of $80\%$. This is probably because the adversarial training of $a$ \textit{vs.} $C$ guides $G$ to generate \textit{hard} counterfactuals, which are efficient to train the classifier. The results demonstrate that our approach could help alleviate \textit{catastrophic forgetting} even with a small number of synthetic samples used for augmentation. This experiment could also be viewed as a measurement of the \textit{sample efficiency}, \textit{i.e.}~how efficient a synthetic sample is in terms of re-training a classifier.

\subsection{Can the proposed procedure alleviate \textit{spurious correlations}?}
\label{sec: can the proposed procedure alleviate spurious correlations}
\textit{Spurious correlation} occurs when two factors appear to be correlated to each other but in fact they are not~\citep{simon1954spurious}. Spurious correlation could affect the performance of deep neural networks and has been actively studied in computer vision field~\citep{sagawa2020investigation,liu2021just,sagawa2020investigation,sagawa2019distributionally,youbi2021simple,goel2021model} and in medical imaging analysis field~\citep{10.3389/fdgth.2021.671015,degrave2021ai}. For instance, suppose we have an dataset of \textit{bird} and \textit{bat} photos. For \textit{bird} photos, most backgrounds are  \textit{sky}. For \textit{bat} photos, most backgrounds are \textit{cave}. If a classifier learns this spurious correlation, \textit{e.g.} it classifies a photo as \textit{bird} as long as the background is \textit{sky}, then it will perform poorly on images where \textit{bats} are flying in the \textit{sky}. In this section, we investigate if our approach could correct such \textit{spurious correlations} by changing $a$ to generate hard counterfactuals.

Here we create a dataset where 7860 images between 60 and 75 yrs old are AD, and 7680 images between 75 and 90 yrs old are healthy, denoted as $D_{spurious}$. This is to construct a \textit{spurious correlation}: $young \rightarrow AD$ and $old \rightarrow CN$ (in reality older people have higher chances of getting AD~\citep{goedert2006century}). Then we pre-train $C$ on $D_{spurious}$. The brain ageing model proposed in \citet{xia2021learning} only considered simulating \textit{ageing} process, {but did not consider brain \textit{rejuvenation}, i.e.,~the reverse of ageing. To utilise old CN data,
% To be able to use \textit{old CN} data,
we pre-train another generator in the \textit{rejuvenation} direction, i.e.,generating \textit{younger} brain images from old ones.}
As a result, we obtain two generators that are pre-trained on $D_{train}$, denoted as $G_{ageing}$ and $G_{rejuve}$, where $G_{rejuve}$ is trained to simulate  the \textit{rejuvenation} process. Fig.~S2 in \textit{Supplementary material} shows visual results of  $G_{rejuve}$. After that, we select 50 CN and 50 AD \textit{hard} images from $D_{spurious}$, denoted as $D_{hard}$ and perform the adversarial classification training using $G_{rejuve}$ for \textit{old CN} samples and $G_{ageing}$ for \textit{young AD} samples. The target ages $a$ are initialized as real ages of $\mathbf{x}$.

\begin{figure}[!t]
    \centering
    \includegraphics[scale=0.65]{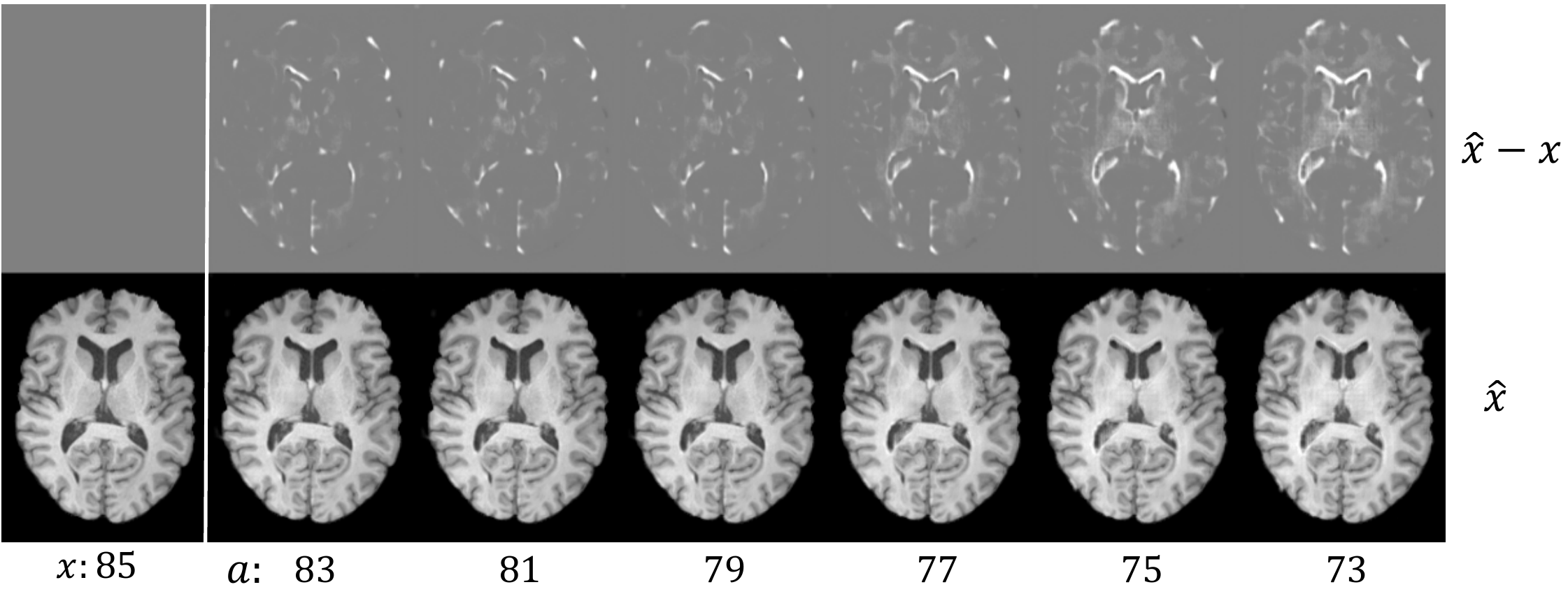}
    \caption{Example results of brain \textit{rejuvenation} for an image ($x$) of a 85 year old CN subject. We synthesise \textit{rejuvenated} images $\hat{x}$ at different target ages $a$. We also show the differences between $\hat{x}$ and $x$, $\hat{x}-x$. For more details see text.}
    \label{fig: brain rejuvenation}
\end{figure}

After we obtain $G_{ageing}$ and $G_{rejuvenation}$, we select 50 CN and 50 AD images from $D_{spurious}$ that result in highest training errors, denoted as $D_{hard}$. Note that the selected CN images are between 75 and 90 yrs old, and the AD images are between 60 and 75 yrs old. Then we generate synthetic images from $D_{hard}$ using $G_{rejuvenation}$ for old CN samples and $G_{ageing}$ for young AD samples. The target ages $a$ are initialized as their ground-truth ages. Finally, we perform the adversarial training between $a$ and the classifier $C$. Here we want to see if the adversarial training can detect the \textit{spurious correlations} purposely created by us, and more importantly, we want to see if the adversarial training between $a$ and $C$ can \textit{break} the spurious correlations. 

Table~\ref{tab: test acc for spurious results} presents the test accuracies of our approach and baselines. For \textit{Na\"ive}, we directly use  the classifier $C$ pre-trained on $D_{spurious}$. For \textit{HSRS}, we randomly generate synthetic samples from $D_{hard}$ for augmentation. For \textit{JTT}, we simply select mis-classified samples from $D_{spurious}$ and up-sample these samples. 

\begin{table}[!b]
\caption{Test accuracies for our procedure and baselines when $C$ pre-trained on $D_{spurious}$. We first present the average test accuracies for different age groups with CN diagnosis (column 2-3) or AD (column 4-5), and then present the average test accuracies for the whole testing set (column 6). For each method, the \textit{worst-group} performance  is shown in \textit{italic}. For each age group, \textit{i.e.}~each column, the \textbf{best} performance was shown in \textbf{bold}. For more details see text. }
\centering
\begin{tabular}{c|cc|cc|c}
\hline
Acc \%   & \multicolumn{2}{c|}{CN}                & \multicolumn{2}{c|}{AD}       &                           \\ \hline
Methods  & 60-75yrs               & 75-90yrs      & 60-75yrs      & 75-90yrs      & Overall                   \\ \hline
Na\"ive    & \textit{40.9}          & 81.6          & \textbf{95.1} & 45.7          & 67.0                      \\ \hline
HSRS     & \textit{60.7}          & 85.3          & 81.1          & 67.2          & 75.0                      \\ \hline
JTT      & {50.5}          & \textbf{88.4} & 85.5          & \textit{40.7}          & 67.9 \\ \hline
proposed & \textit{\textbf{73.1}} & 83.4          & 81.5          & \textbf{75.8} & \textbf{79.0}             \\ \hline
\end{tabular}
\label{tab: test acc for spurious results}
\end{table}

We can observe from Table~\ref{tab: test acc for spurious results} that the pre-trained $C$ on $D_{spurious}$ (\textit{Na\"ive}) achieves much worse performance ($67.0\%$ accuracy) compared to that of Table~\ref{tab: accuracy for main experiment} ($88.4\%$ accuracy). Specifically, it tends to misclassify \textit{young CN} images as AD and misclassify \textit{old AD} images as CN. This is likely due to the spurious correlations that we purposely create in $D_{spurious}$: $young \rightarrow AD$ and $old \rightarrow CN$. We notice that for \textit{Na\"ive}, the test accuracies of AD groups are higher than that of CN groups. This is likely due to the fact we have more AD training data, and the classifier is biased to classify a subject to AD. This can be viewed as another \textit{spurious correlation}. Overall, we observe that our method achieves the best results, followed by \textit{HSRS}. This shows that the synthetic results generated by the generators are helpful to alleviate the effect of \textit{spurious correlations} and improve downstream tasks. The improvement of our approach over \textit{HSRS} is due to the adversarial training between $a$ and $C$, which guides the generator to produce hard counterfactuals. We observe \textit{JTT} does not improve the test accuracies significantly. A potential reason is that \textit{JTT} tries to find `hard' samples in the training dataset. However, in this experiment, the `hard' samples should be \textit{young CN} and \textit{old AD }samples which do not exist in the training dataset $D_{spurious}$. By contrast, our procedure could guide $G$ to generate these samples, and \textit{HSRS} could create these samples by random chance.

\begin{figure}[tb]
    \centering
    \includegraphics[scale=0.56]{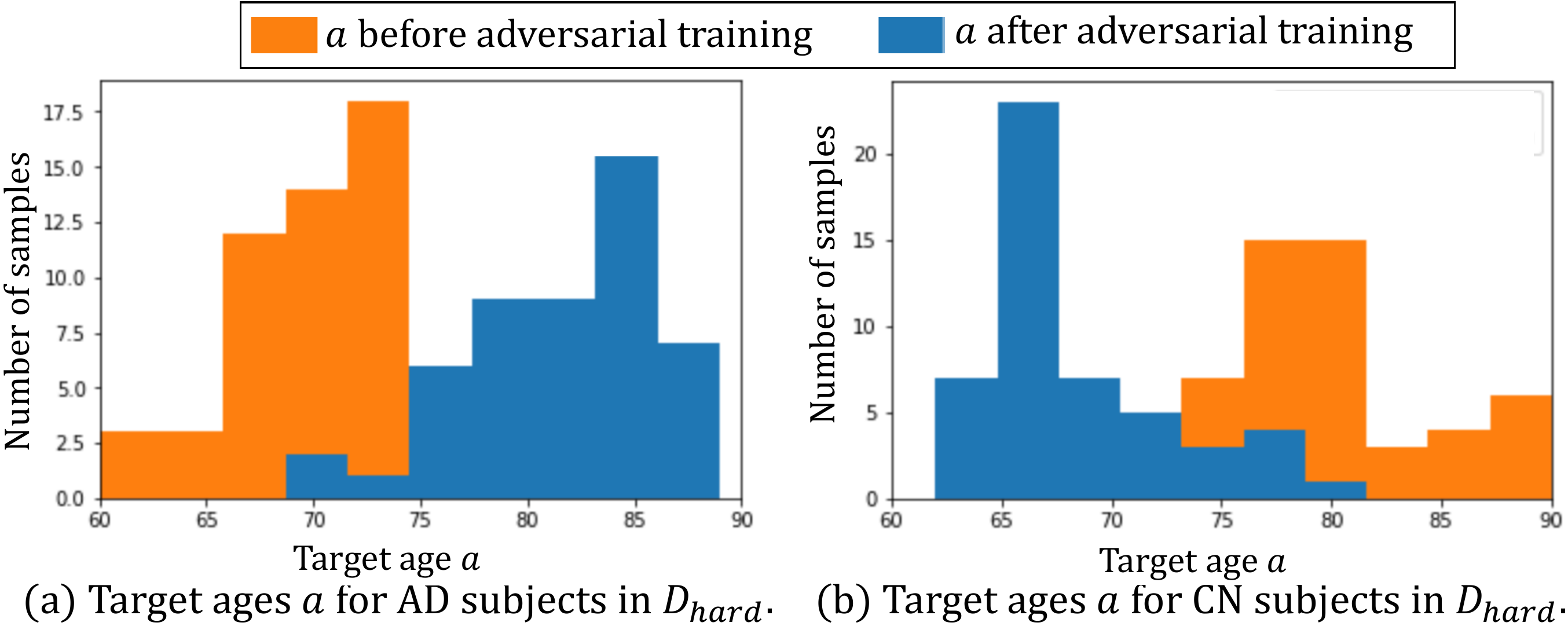}
    \caption{Histograms of target ages $a$ before and after adversarial training: (a) the histogram of $a$ for the 50 AD subjects in $D_{hard}$; (b) the histogram of $a$ for the 50 CN subjects in $D_{hard}$. Here we show histograms of $a$ before (in orange) and after (in blue) the adversarial training.  }
    \label{fig: histograms of a}
\end{figure}

Figure~\ref{fig: histograms of a}  plots the histograms of the target ages $a$ before and after the adversarial training. From Figure~\ref{fig: histograms of a} we can observe that the adversarial training pushes $a$ towards the \textit{hard} direction, which could alleviate the spurious correlations. For instance, in $D_{spurious}$ and $D_{hard}$ the AD subjects are all in the \textit{young} group, \textit{i.e.}~60-75 yrs old, and the classifier learns the \textit{spurious correlation}: $young \rightarrow AD$, but in Figure~\ref{fig: histograms of a} (a) we can observe that the adversarial training learns to generate  AD synthetic images in the range of 75-90 yrs old. These \textit{old AD} synthetic images can help alleviate the spurious correlation and improve the performance of $C$. Similarly, we can observe $a$ are pushed towards \textit{young} for CN subjects in Figure~\ref{fig: histograms of a} (b).

\section{Conclusion}
\label{sec: chap 6 conclusion}
We presented a novel adversarial counterfactual scheme to utilise conditional generative models for downstream tasks, e.g. classification. The proposed procedure formulates an adversarial game between the \textit{conditional factor} of a pre-trained generative model and the downstream \textit{classifier}.
We improved a recent brain ageing synthesis model as the generative model and focused on AD classification. We presented quantitative results showing that our approach can improve the performance of the AD classifier. Furthermore, we showed that our method could alleviate the effect of \textit{catastrophic forgetting} and \textit{spurious correlations}. There are several future directions. The proposed approach can be potentially applied to other generators on other datasets. The way we updated the conditional factor (target age) could be improved. Instead of a continuous scalar (target age), we can consider extending the proposed adversarial counterfactual augmentation to update other types of  conditional factors, e.g., discrete factor or image. {The strategy that we used to select \textit{hard} samples may not be the most effective and could be improved}.

\section*{Conflict of Interest Statement}

\textbf{The authors declare that the research was conducted in the absence of any commercial or financial relationships that could be construed as a potential conflict of interest.}

\section*{Author Contributions}

TX, PS, CQ and SAT contributed to the conceptualization of this work. TX, PS, CQ and SAT designed the methodology. TX developed the software tools necessary for preprocessing and analysing images files and for training the model. TX drafted this manuscript. All authors reviewed the manuscript. 

% \section*{Funding}
%  {++++++++To be added++++++++++++}

\section*{Acknowledgments}
This work was supported by the University of Edinburgh, the Royal Academy of Engineering and Canon Medical Research Europe via PhD studentships of Pedro Sanchez (grant RCSRF1819/825). This work was partially supported by the Alan Turing Institute under the EPSRC grant EP//N510129//1. S.A. Tsaftaris acknowledges the support of Canon Medical and the Royal Academy of Engineering and the Research Chairs and Senior Research Fellowships scheme (grant RCSRF1819/8/25).

% \section*{Supplemental Data}
%  \href{http://home.frontiersin.org/about/author-guidelines#SupplementaryMaterial}{Supplementary Material} should be uploaded separately on submission, if there are Supplementary Figures, please include the caption in the same file as the figure. LaTeX Supplementary Material templates can be found in the Frontiers LaTeX folder.

% \section*{Data Availability Statement}
% The datasets [GENERATED/ANALYZED] for this study can be found in the [NAME OF REPOSITORY] [LINK].
% Please see the availability of data guidelines for more information, at https://www.frontiersin.org/about/author-guidelines#AvailabilityofData

\bibliographystyle{Frontiers-Harvard} %  Many Frontiers journals use the Harvard referencing system (Author-date), to find the style and resources for the journal you are submitting to: https://zendesk.frontiersin.org/hc/en-us/articles/360017860337-Frontiers-Reference-Styles-by-Journal. For Humanities and Social Sciences articles please include page numbers in the in-text citations 
\bibliography{test}

\begin{thebibliography}{52}
\providecommand{\natexlab}[1]{#1}
\expandafter\ifx\csname urlstyle\endcsname\relax
  \providecommand{\doi}[1]{doi:\discretionary{}{}{}#1}\else
  \providecommand{\doi}{doi:\discretionary{}{}{}\begingroup
  \urlstyle{rm}\Url}\fi
\providecommand{\selectlanguage}[1]{\relax}
\providecommand{\bibAnnoteFile}[1]{%
  \IfFileExists{#1}{\begin{quotation}\noindent\textsc{Key:} #1\\
  \textsc{Annotation:}\ \input{#1}\end{quotation}}{}}
\providecommand{\bibAnnote}[2]{%
  \begin{quotation}\noindent\textsc{Key:} #1\\
  \textsc{Annotation:}\ #2\end{quotation}}

\bibitem[{Aljundi et~al.(2017)Aljundi, Chakravarty, and
  Tuytelaars}]{aljundi2017expert}
Aljundi, R., Chakravarty, P., and Tuytelaars, T. (2017).
\newblock Expert gate: Lifelong learning with a network of experts.
\newblock In \emph{Proceedings of the IEEE Conference on Computer Vision and
  Pattern Recognition}. 3366--3375
\bibAnnoteFile{aljundi2017expert}

\bibitem[{Aljundi et~al.(2018)Aljundi, Rohrbach, and
  Tuytelaars}]{aljundi2018selfless}
Aljundi, R., Rohrbach, M., and Tuytelaars, T. (2018).
\newblock Selfless sequential learning.
\newblock In \emph{International Conference on Learning Representations}
\bibAnnoteFile{aljundi2018selfless}

\bibitem[{Antoniou et~al.(2018)Antoniou, Storkey, and
  Edwards}]{antoniou2017data}
Antoniou, A., Storkey, A., and Edwards, H. (2018).
\newblock Data augmentation generative adversarial networks.
\newblock \emph{arXiv preprint arXiv:1711.04340}
\bibAnnoteFile{antoniou2017data}

\bibitem[{Bowles et~al.(2018)Bowles, Chen, Guerrero, Bentley, Gunn, Hammers
  et~al.}]{bowles2018gan}
Bowles, C., Chen, L., Guerrero, R., Bentley, P., Gunn, R., Hammers, A., et~al.
  (2018).
\newblock Gan augmentation: Augmenting training data using generative
  adversarial networks.
\newblock \emph{arXiv preprint arXiv:1810.10863}
\bibAnnoteFile{bowles2018gan}

\bibitem[{Chaudhry et~al.(2018)Chaudhry, Dokania, Ajanthan, and
  Torr}]{chaudhry2018riemannian}
Chaudhry, A., Dokania, P.~K., Ajanthan, T., and Torr, P.~H. (2018).
\newblock Riemannian walk for incremental learning: Understanding forgetting
  and intransigence.
\newblock In \emph{Proceedings of the European Conference on Computer Vision
  (ECCV)}. 532--547
\bibAnnoteFile{chaudhry2018riemannian}

\bibitem[{Chaudhry et~al.(2019)Chaudhry, Rohrbach, Elhoseiny, Ajanthan,
  Dokania, Torr et~al.}]{DBLP:journals/corr/abs-1902-10486}
Chaudhry, A., Rohrbach, M., Elhoseiny, M., Ajanthan, T., Dokania, P.~K., Torr,
  P. H.~S., et~al. (2019).
\newblock Continual learning with tiny episodic memories.
\newblock \emph{CoRR} abs/1902.10486
\bibAnnoteFile{DBLP:journals/corr/abs-1902-10486}

\bibitem[{Chen et~al.(2021)Chen, Qin, Ouyang, Wang, Qiu, Chen
  et~al.}]{chen2021enhancing}
Chen, C., Qin, C., Ouyang, C., Wang, S., Qiu, H., Chen, L., et~al. (2021).
\newblock Enhancing mr image segmentation with realistic adversarial data
  augmentation.
\newblock \emph{arXiv preprint arXiv:2108.03429}
\bibAnnoteFile{chen2021enhancing}

\bibitem[{Chen and Su(2021)}]{chen2021sample}
Chen, J. and Su, B. (2021).
\newblock Sample-specific and context-aware augmentation for long tail image
  classification
\bibAnnoteFile{chen2021sample}

\bibitem[{Chen and Liu(2018)}]{chen2018lifelong}
Chen, Z. and Liu, B. (2018).
\newblock Lifelong machine learning.
\newblock \emph{Synthesis Lectures on Artificial Intelligence and Machine
  Learning} 12, 1--207
\bibAnnoteFile{chen2018lifelong}

\bibitem[{Chlap et~al.(2021)Chlap, Min, Vandenberg, Dowling, Holloway, and
  Haworth}]{chlap2021review}
Chlap, P., Min, H., Vandenberg, N., Dowling, J., Holloway, L., and Haworth, A.
  (2021).
\newblock A review of medical image data augmentation techniques for deep
  learning applications.
\newblock \emph{Journal of Medical Imaging and Radiation Oncology} 65, 545--563
\bibAnnoteFile{chlap2021review}

\bibitem[{Cubuk et~al.(2019)Cubuk, Zoph, Mane, Vasudevan, and
  Le}]{cubuk2019autoaugment}
Cubuk, E.~D., Zoph, B., Mane, D., Vasudevan, V., and Le, Q.~V. (2019).
\newblock Autoaugment: Learning augmentation strategies from data.
\newblock In \emph{Proceedings of the IEEE/CVF Conference on Computer Vision
  and Pattern Recognition}. 113--123
\bibAnnoteFile{cubuk2019autoaugment}

\bibitem[{Dar et~al.(2019)Dar, Yurt, Karacan, Erdem, Erdem, and
  {\c{C}}ukur}]{dar2019image}
Dar, S.~U., Yurt, M., Karacan, L., Erdem, A., Erdem, E., and {\c{C}}ukur, T.
  (2019).
\newblock Image synthesis in multi-contrast mri with conditional generative
  adversarial networks.
\newblock \emph{IEEE transactions on medical imaging}
\bibAnnoteFile{dar2019image}

\bibitem[{Dash et~al.(2022)Dash, Balasubramanian, and
  Sharma}]{dash2022evaluating}
Dash, S., Balasubramanian, V.~N., and Sharma, A. (2022).
\newblock Evaluating and mitigating bias in image classifiers: A causal
  perspective using counterfactuals.
\newblock In \emph{Proceedings of the IEEE/CVF Winter Conference on
  Applications of Computer Vision}. 915--924
\bibAnnoteFile{dash2022evaluating}

\bibitem[{DeGrave et~al.(2021)DeGrave, Janizek, and Lee}]{degrave2021ai}
DeGrave, A.~J., Janizek, J.~D., and Lee, S.-I. (2021).
\newblock Ai for radiographic covid-19 detection selects shortcuts over signal.
\newblock \emph{Nature Machine Intelligence} , 1--10
\bibAnnoteFile{degrave2021ai}

\bibitem[{Delange et~al.(2021)Delange, Aljundi, Masana, Parisot, Jia, Leonardis
  et~al.}]{delange2021continual}
Delange, M., Aljundi, R., Masana, M., Parisot, S., Jia, X., Leonardis, A.,
  et~al. (2021).
\newblock A continual learning survey: Defying forgetting in classification
  tasks.
\newblock \emph{IEEE Transactions on Pattern Analysis and Machine Intelligence}
\bibAnnoteFile{delange2021continual}

\bibitem[{Dietterich(1995)}]{dietterich1995overfitting}
Dietterich, T. (1995).
\newblock Overfitting and undercomputing in machine learning.
\newblock \emph{ACM computing surveys (CSUR)} 27, 326--327
\bibAnnoteFile{dietterich1995overfitting}

\bibitem[{Feldman and Zhang(2020)}]{feldman2020neural}
Feldman, V. and Zhang, C. (2020).
\newblock What neural networks memorize and why: Discovering the long tail via
  influence estimation.
\newblock \emph{arXiv preprint arXiv:2008.03703}
\bibAnnoteFile{feldman2020neural}

\bibitem[{Frid-Adar et~al.(2018{\natexlab{a}})Frid-Adar, Diamant, Klang,
  Amitai, Goldberger, and Greenspan}]{frid2018gan}
Frid-Adar, M., Diamant, I., Klang, E., Amitai, M., Goldberger, J., and
  Greenspan, H. (2018{\natexlab{a}}).
\newblock Gan-based synthetic medical image augmentation for increased cnn
  performance in liver lesion classification.
\newblock \emph{Neurocomputing} 321, 321--331
\bibAnnoteFile{frid2018gan}

\bibitem[{Frid-Adar et~al.(2018{\natexlab{b}})Frid-Adar, Klang, Amitai,
  Goldberger, and Greenspan}]{frid2018synthetic}
Frid-Adar, M., Klang, E., Amitai, M., Goldberger, J., and Greenspan, H.
  (2018{\natexlab{b}}).
\newblock Synthetic data augmentation using gan for improved liver lesion
  classification.
\newblock In \emph{2018 IEEE 15th international symposium on biomedical imaging
  (ISBI 2018)} (IEEE), 289--293
\bibAnnoteFile{frid2018synthetic}

\bibitem[{Gao et~al.(2021)Gao, Tang, Zhou, and Metaxas}]{gao2021enabling}
Gao, Y., Tang, Z., Zhou, M., and Metaxas, D. (2021).
\newblock Enabling data diversity: Efficient automatic augmentation via
  regularized adversarial training.
\newblock In \emph{International Conference on Information Processing in
  Medical Imaging} (Springer), 85--97
\bibAnnoteFile{gao2021enabling}

\bibitem[{Gepperth and Karaoguz(2016)}]{gepperth2016bio}
Gepperth, A. and Karaoguz, C. (2016).
\newblock A bio-inspired incremental learning architecture for applied
  perceptual problems.
\newblock \emph{Cognitive Computation} 8, 924--934
\bibAnnoteFile{gepperth2016bio}

\bibitem[{Goedert and Spillantini(2006)}]{goedert2006century}
Goedert, M. and Spillantini, M.~G. (2006).
\newblock A century of alzheimer's disease.
\newblock \emph{Science} 314, 777--781
\bibAnnoteFile{goedert2006century}

\bibitem[{Goel et~al.(2021)Goel, Gu, Li, and R{\'e}}]{goel2021model}
Goel, K., Gu, A., Li, Y., and R{\'e}, C. (2021).
\newblock Model patching: Closing the subgroup performance gap with data
  augmentation.
\newblock \emph{ICLR}
\bibAnnoteFile{goel2021model}

\bibitem[{Gong et~al.(2021{\natexlab{a}})Gong, Ren, Ye, and
  Liu}]{gong2021maxup}
Gong, C., Ren, T., Ye, M., and Liu, Q. (2021{\natexlab{a}}).
\newblock Maxup: Lightweight adversarial training with data augmentation
  improves neural network training.
\newblock In \emph{Proceedings of the IEEE/CVF Conference on Computer Vision
  and Pattern Recognition}. 2474--2483
\bibAnnoteFile{gong2021maxup}

\bibitem[{Gong et~al.(2021{\natexlab{b}})Gong, Wang, Li, Chandra, and
  Liu}]{gong2021keepaugment}
Gong, C., Wang, D., Li, M., Chandra, V., and Liu, Q. (2021{\natexlab{b}}).
\newblock Keepaugment: A simple information-preserving data augmentation
  approach.
\newblock In \emph{Proceedings of the IEEE/CVF Conference on Computer Vision
  and Pattern Recognition}. 1055--1064
\bibAnnoteFile{gong2021keepaugment}

\bibitem[{Kirkpatrick et~al.(2017)Kirkpatrick, Pascanu, Rabinowitz, Veness,
  Desjardins, Rusu et~al.}]{kirkpatrick2017overcoming}
Kirkpatrick, J., Pascanu, R., Rabinowitz, N., Veness, J., Desjardins, G., Rusu,
  A.~A., et~al. (2017).
\newblock Overcoming catastrophic forgetting in neural networks.
\newblock \emph{Proceedings of the national academy of sciences} 114,
  3521--3526
\bibAnnoteFile{kirkpatrick2017overcoming}

\bibitem[{Li et~al.(2021)Li, Gong, Liu, Wang, Qiao, and Cheng}]{li2021metasaug}
Li, S., Gong, K., Liu, C.~H., Wang, Y., Qiao, F., and Cheng, X. (2021).
\newblock Metasaug: Meta semantic augmentation for long-tailed visual
  recognition.
\newblock In \emph{Proceedings of the IEEE/CVF Conference on Computer Vision
  and Pattern Recognition}. 5212--5221
\bibAnnoteFile{li2021metasaug}

\bibitem[{Liu et~al.(2021)Liu, Haghgoo, Chen, Raghunathan, Koh, Sagawa
  et~al.}]{liu2021just}
Liu, E.~Z., Haghgoo, B., Chen, A.~S., Raghunathan, A., Koh, P.~W., Sagawa, S.,
  et~al. (2021).
\newblock Just train twice: Improving group robustness without training group
  information.
\newblock In \emph{International Conference on Machine Learning} (PMLR),
  6781--6792
\bibAnnoteFile{liu2021just}

\bibitem[{Lopez-Paz and Ranzato(2017)}]{lopez2017gradient}
Lopez-Paz, D. and Ranzato, M. (2017).
\newblock Gradient episodic memory for continual learning.
\newblock \emph{Advances in neural information processing systems} 30,
  6467--6476
\bibAnnoteFile{lopez2017gradient}

\bibitem[{Mahmood et~al.(2021)Mahmood, Shrestha, Bates, Mannelli, Corrias, Erdi
  et~al.}]{10.3389/fdgth.2021.671015}
Mahmood, U., Shrestha, R., Bates, D. D.~B., Mannelli, L., Corrias, G., Erdi,
  Y.~E., et~al. (2021).
\newblock Detecting spurious correlations with sanity tests for artificial
  intelligence guided radiology systems.
\newblock \emph{Frontiers in Digital Health} 3
\bibAnnoteFile{10.3389/fdgth.2021.671015}

\bibitem[{McCloskey and Cohen(1989)}]{mccloskey1989catastrophic}
McCloskey, M. and Cohen, N.~J. (1989).
\newblock Catastrophic interference in connectionist networks: The sequential
  learning problem.
\newblock In \emph{Psychology of learning and motivation} (Elsevier), vol.~24.
  109--165
\bibAnnoteFile{mccloskey1989catastrophic}

\bibitem[{Mildenhall et~al.(2020)Mildenhall, Srinivasan, Tancik, Barron,
  Ramamoorthi, and Ng}]{mildenhall2020nerf}
Mildenhall, B., Srinivasan, P.~P., Tancik, M., Barron, J.~T., Ramamoorthi, R.,
  and Ng, R. (2020).
\newblock Nerf: Representing scenes as neural radiance fields for view
  synthesis.
\newblock In \emph{European conference on computer vision} (Springer), 405--421
\bibAnnoteFile{mildenhall2020nerf}

\bibitem[{Oh et~al.(2021)Oh, Yoon, and Suk}]{oh2021learn}
Oh, K., Yoon, J.~S., and Suk, H.-I. (2021).
\newblock Learn-explain-reinforce: Counterfactual reasoning and its guidance to
  reinforce an alzheimer's disease diagnosis model.
\newblock \emph{arXiv preprint arXiv:2108.09451}
\bibAnnoteFile{oh2021learn}

\bibitem[{Parisi et~al.(2019)Parisi, Kemker, Part, Kanan, and
  Wermter}]{parisi2019continual}
Parisi, G.~I., Kemker, R., Part, J.~L., Kanan, C., and Wermter, S. (2019).
\newblock Continual lifelong learning with neural networks: A review.
\newblock \emph{Neural Networks} 113, 54--71
\bibAnnoteFile{parisi2019continual}

\bibitem[{Petersen et~al.(2010)Petersen, Aisen, Beckett, Donohue, Gamst, Harvey
  et~al.}]{petersen2010alzheimer}
Petersen, R.~C., Aisen, P., Beckett, L.~A., Donohue, M., Gamst, A., Harvey,
  D.~J., et~al. (2010).
\newblock {Alzheimer's disease neuroimaging initiative (ADNI): clinical
  characterization}.
\newblock \emph{Neurology} 74, 201--209
\bibAnnoteFile{petersen2010alzheimer}

\bibitem[{Robins(1995)}]{robins1995catastrophic}
Robins, A. (1995).
\newblock Catastrophic forgetting, rehearsal and pseudorehearsal.
\newblock \emph{Connection Science} 7, 123--146
\bibAnnoteFile{robins1995catastrophic}

\bibitem[{Sagawa et~al.(2019)Sagawa, Koh, Hashimoto, and
  Liang}]{sagawa2019distributionally}
Sagawa, S., Koh, P.~W., Hashimoto, T.~B., and Liang, P. (2019).
\newblock Distributionally robust neural networks for group shifts: On the
  importance of regularization for worst-case generalization.
\newblock \emph{arXiv preprint arXiv:1911.08731}
\bibAnnoteFile{sagawa2019distributionally}

\bibitem[{Sagawa et~al.(2020)Sagawa, Raghunathan, Koh, and
  Liang}]{sagawa2020investigation}
Sagawa, S., Raghunathan, A., Koh, P.~W., and Liang, P. (2020).
\newblock An investigation of why overparameterization exacerbates spurious
  correlations.
\newblock In \emph{International Conference on Machine Learning} (PMLR),
  8346--8356
\bibAnnoteFile{sagawa2020investigation}

\bibitem[{Shamsolmoali et~al.(2021)Shamsolmoali, Zareapoor, Shen, Sadka, and
  Yang}]{shamsolmoali2021imbalanced}
Shamsolmoali, P., Zareapoor, M., Shen, L., Sadka, A.~H., and Yang, J. (2021).
\newblock Imbalanced data learning by minority class augmentation using capsule
  adversarial networks.
\newblock \emph{Neurocomputing} 459, 481--493
\bibAnnoteFile{shamsolmoali2021imbalanced}

\bibitem[{Shen et~al.(2017)Shen, Wu, and Suk}]{shen2017deep}
Shen, D., Wu, G., and Suk, H.-I. (2017).
\newblock Deep learning in medical image analysis.
\newblock \emph{Annual review of biomedical engineering} 19, 221--248
\bibAnnoteFile{shen2017deep}

\bibitem[{Shin et~al.(2018)Shin, Tenenholtz, Rogers, Schwarz, Senjem, Gunter
  et~al.}]{shin2018medical}
Shin, H.-C., Tenenholtz, N.~A., Rogers, J.~K., Schwarz, C.~G., Senjem, M.~L.,
  Gunter, J.~L., et~al. (2018).
\newblock {Medical Image Synthesis for Data Augmentation and Anonymization
  Using Generative Adversarial Networks}.
\newblock In \emph{International Workshop on Simulation and Synthesis in
  Medical Imaging} (Springer), 1--11
\bibAnnoteFile{shin2018medical}

\bibitem[{Shorten and Khoshgoftaar(2019)}]{shorten2019survey}
Shorten, C. and Khoshgoftaar, T.~M. (2019).
\newblock A survey on image data augmentation for deep learning.
\newblock \emph{Journal of Big Data} 6, 1--48
\bibAnnoteFile{shorten2019survey}

\bibitem[{Simon(1954)}]{simon1954spurious}
Simon, H.~A. (1954).
\newblock Spurious correlation: A causal interpretation.
\newblock \emph{Journal of the American statistical Association} 49, 467--479
\bibAnnoteFile{simon1954spurious}

\bibitem[{Simonyan and Zisserman(2015)}]{simonyan2015very}
Simonyan, K. and Zisserman, A. (2015).
\newblock {Very deep convolutional networks for large-scale image recognition}.
\newblock \emph{ICLR}
\bibAnnoteFile{simonyan2015very}

\bibitem[{Srivastava et~al.(2014)Srivastava, Hinton, Krizhevsky, Sutskever, and
  Salakhutdinov}]{srivastava2014dropout}
Srivastava, N., Hinton, G., Krizhevsky, A., Sutskever, I., and Salakhutdinov,
  R. (2014).
\newblock Dropout: a simple way to prevent neural networks from overfitting.
\newblock \emph{The journal of machine learning research} 15, 1929--1958
\bibAnnoteFile{srivastava2014dropout}

\bibitem[{Tancik et~al.(2020)Tancik, Srinivasan, Mildenhall, Fridovich-Keil,
  Raghavan, Singhal et~al.}]{tancik2020fourier}
Tancik, M., Srinivasan, P.~P., Mildenhall, B., Fridovich-Keil, S., Raghavan,
  N., Singhal, U., et~al. (2020).
\newblock Fourier features let networks learn high frequency functions in low
  dimensional domains.
\newblock \emph{NeurIPS}
\bibAnnoteFile{tancik2020fourier}

\bibitem[{van~de Ven et~al.(2020)van~de Ven, Siegelmann, and
  Tolias}]{van2020brain}
van~de Ven, G.~M., Siegelmann, H.~T., and Tolias, A.~S. (2020).
\newblock Brain-inspired replay for continual learning with artificial neural
  networks.
\newblock \emph{Nature communications} 11, 1--14
\bibAnnoteFile{van2020brain}

\bibitem[{Wang et~al.(2021)Wang, Huang, Song, Pan, Xia, and
  Wu}]{wang2021regularizing}
Wang, Y., Huang, G., Song, S., Pan, X., Xia, Y., and Wu, C. (2021).
\newblock Regularizing deep networks with semantic data augmentation.
\newblock \emph{IEEE Transactions on Pattern Analysis and Machine Intelligence}
\bibAnnoteFile{wang2021regularizing}

\bibitem[{Woolrich et~al.(2009)Woolrich, Jbabdi, Patenaude, Chappell, Makni,
  Behrens et~al.}]{woolrich2009bayesian}
Woolrich, M.~W., Jbabdi, S., Patenaude, B., Chappell, M., Makni, S., Behrens,
  T., et~al. (2009).
\newblock Bayesian analysis of neuroimaging data in {FSL}.
\newblock \emph{Neuroimage} 45, S173--S186
\bibAnnoteFile{woolrich2009bayesian}

\bibitem[{Xia et~al.(2021)Xia, Chartsias, Wang, Tsaftaris, Initiative
  et~al.}]{xia2021learning}
Xia, T., Chartsias, A., Wang, C., Tsaftaris, S.~A., Initiative, A. D.~N.,
  et~al. (2021).
\newblock Learning to synthesise the ageing brain without longitudinal data.
\newblock \emph{Medical Image Analysis} 73, 102169
\bibAnnoteFile{xia2021learning}

\bibitem[{Youbi~Idrissi et~al.(2021)Youbi~Idrissi, Arjovsky, Pezeshki, and
  Lopez-Paz}]{youbi2021simple}
Youbi~Idrissi, B., Arjovsky, M., Pezeshki, M., and Lopez-Paz, D. (2021).
\newblock Simple data balancing achieves competitive worst-group-accuracy.
\newblock \emph{arXiv e-prints} , arXiv--2110
\bibAnnoteFile{youbi2021simple}

\bibitem[{Zhang et~al.(2020)Zhang, Wang, Liu, Lin, and Ling}]{zhang2020deep}
Zhang, X., Wang, Z., Liu, D., Lin, Q., and Ling, Q. (2020).
\newblock Deep adversarial data augmentation for extremely low data regimes.
\newblock \emph{IEEE Transactions on Circuits and Systems for Video Technology}
  31, 15--28
\bibAnnoteFile{zhang2020deep}

\end{thebibliography}

\end{document}